 \documentclass[12pt,preprint]{aastex}  

\slugcomment{ApJ, Received 05/05/2013, Accepted 08/02/2013}

\usepackage{natbib,hyperref,CJK}

\shorttitle{Young Open Cluster G144.9$+$0.4}
\shortauthors{Lin, Chen, \& Panwar}

\begin{document}

\title{Characterization of the Young Open Cluster G144.9$+$0.4 in the 
Camelopardalis OB1 Association}

\begin{CJK*}{UTF8}{bkai}
\author{C. C. Lin (林建爭)}
\author{W. P. Chen (陳文屏)}
\author{N. Panwar}
\affil{Graduate Institute of Astronomy, National Central University, 
300 Jhongda Rd., Jhongli 32001, Taiwan; cclin@astro.ncu.edu.tw}

\begin{abstract}
Our star-count analysis of the Two Micron All Sky Survey point sources resulted 
in an identification of the star cluster G144.9$+$0.4.  The cluster was found, 
but not characterized, by \citeauthor{glu10}  We show that the cluster is 
physically associated with the Cam\,OB1 association at a distance of about 
1~kpc and with an age of 1--2~Myr.  Pre-main sequence stars are identified, on 
the basis of photometric and proper motion data.  A total of 91 additional OB 
star candidates was found in subgroups 1A and 1B, a significant increase from 
the currently known 43 OB stars.  The OB members show an age spread that 
indicates a sustained star formation for at least the last 10--15 Myr.  The 
young cluster G144.9$+$0.4 represents the latest episode of sequential star 
formation in this cloud complex. 
\end{abstract}


\section{Introduction}

Hundreds of thousands of open clusters (OCs) should currently exist in the 
Milky Way galaxy based on the number of OCs present in the solar neighborhood 
\citep{pis06}.  However, the databases of OCs \citep{dnb01, dia02, bic03, 
dut03, kro06, fro07} contain only a few thousand entries that are limited to 
OCs within 1~kpc.  This discrepancy is due partly to dust extinction in the 
Galactic plane, and partly to a lack of comprehensive all-sky searches for 
distant systems.  

Star clusters are groupings of member stars in a six-dimensional phase space 
in position and motion.  Kinematic studies of star clusters require special 
instrumentation and are often time-consuming.  Initial identification of a 
star cluster via space grouping, i.e., by the ``star-count'' technique is 
relatively straightforward and has been exploited efficiently on wide-field or 
all-sky surveys \citep{sch11}.  The Two Micron All Sky Survey point source 
catalog \citep[2MASS;][]{cut03} provides a uniformly calibrated database of 
the entire sky in the near-infrared (NIR) wavebands, which allows us to 
recognize OCs even with moderate dust extinction, i.e., partially embedded, 
young star clusters. Earlier works by \cite{bic03}, \cite{dut03}, and 
\cite{fro07} have indeed found hundreds of previously unknown infrared 
clusters using the 2MASS catalog.  

We have developed a star-count technique to recognize star density 
enhancements and have applied it to the sky within $|b|<50\degr$ using the 
2MASS data.  Hundreds of density peaks were found.  We then used the SIMBAD 
database to match these peaks with known sources within a radius of 5\arcmin, 
resulting in 501 OCs, 89 globular clusters, 35 galaxies, 55 galaxy clusters, 
11 H\,{\scriptsize II} regions, and 24 regions contaminated by neighboring 
bright stars.  A total of 5 candidates remain unaccounted for.  One such 
density peak is G144.9$+$0.4 (designated as G144) located at 
$\ell\approx 144\fdg9$ and $b\approx 0\fdg4$, or 
R.A. (J2000)$=03^{\mathrm{h}}39^{\mathrm{m}}16\fs7$, 
decl. (J2000) $=+55\degr58\arcmin24\arcsec$, 
seen toward the Camelopardalis (Cam) OB1 association \citep{lin12}.  This 
cluster candidate was recognized by \citet{glu10}, who used a pipeline 
\citep{kop08} to search for star clusters with 2MASS data.  However, the 
cluster candidate was not well characterized by their pipeline, apparently 
because of the nebulous contamination.  The full results of our OC finding 
will be reported elsewhere.  Here we present a characterization of this cluster 
using photometric and proper motion (PM) measurements, and discuss the context 
of the cluster amid the OB association and cloud complex.  

\section{Photometric, Kinematic, and Spectroscopic Data}
Our data consist of the 2MASS and {\it WISE} infrared magnitudes, PPMXL PMs, 
and our own photometric and spectroscopic observations.  
\vspace{.1in}\\
{\it 2MASS.}  
The NIR magnitudes for point sources have been obtained from 2MASS to the 
10$\sigma$ limiting magnitudes of $J$~(1.25~\micron)$=15.8$, $H$~(1.65~\micron)
$=15.1$, and $K$~(2.17~\micron)$=14.3$, respectively.  For our star count 
analysis, we included only stars with a photometric quality flag better than 
``CCC'', which effectively gives a signal-to-noise ratio $(\mbox{S/N})\ge5$ 
and a photometric uncertainty 
of $<0.2$~mag in every band to ensure completeness.  Subsequently, to 
characterize clusters we constrained only stars with a photometric quality 
flag of ``AAA'', which corresponds to a $\mbox{S/N}\ge10$ and a photometric 
uncertainty of $<0.1$~mag in every band for photometric accuracy.
\vspace{.1in}\\
{\it WISE.}
The {\it Wide-field Infrared Survey Explorer} mapped the whole sky at 3.4, 
4.6, 12, and 22~\micron\ (designated as $W1$, $W2$, $W3$, $W4$) with an 
angular resolution of 6\farcs1, 6\farcs4, 6\farcs5, and 12\farcs0, and a 
5$\sigma$ point-source sensitivity of 0.08, 0.11, 1 and 6~mJy, respectively, 
corresponding to Vega magnitudes of $W1 = 16.6$, $W2 = 15.6$, $W3 = 11.3$, 
and $W4 = 8.0$ \citep[{\it WISE};][]{wri10}.  We exploited the identification 
scheme to search for potential young stellar objects (YSOs) used by 
\citet{koe12}.  The {\it WISE} point sources within the cluster region are 
extracted from Vizier.\footnote{\url{http://vizier.u-strasbg.fr/}}  Only 
sources having 2MASS counterparts are included in the analysis.  
\vspace{.1in}\\
{\it PPMXL.}
In addition to photometric data, kinematic information is also employed to 
secure the membership in a star cluster.  The PPMXL Catalog \citep{ros10} 
contains stars down to about $V \approx 20$~mag and lists the USNO-B1.0 PMs 
\citep{mon03} and 2MASS $JHK$ magnitudes.  The stars with PM uncertainties 
$> 5.0$~mas~yr$^{-1}$ have not been used in our analysis.
\vspace{.1in}\\
{\it Tenagra.}
Narrow-band imaging observations were carried out using the Tenagra II 
0.81~meter $f$/7 telescope in Arizona on 2012 March 30.  The CCD camera used 
an SITe chip with $1024\times1024$ pixels and a pixel size of 
$\sim$0\farcs8, rendering a field of view of $\sim 15\times15$~arcmin$^2$ on 
the sky.  The total exposure times were 540~s for the H$\alpha$ and $R$ bands, 
and 300~s for the $I$ band.  All images were reduced with standard routines, 
including dark and bias subtraction and flat-field normalization.  
SExtractor\footnote{\url{http://www.astromatic.net/software/sextractor}} was 
used for photometry.  The stellar astrometric solution calibrated with the 
2MASS coordinates was calculated with the astrometry package from 
Astrometry.net.\footnote{\url{http://astrometry.net/}}   The Tenagra data 
reached about $R\sim18$~mag.  Supplementary H$\alpha$ data for fainter stars 
were taken from the INT/WFC Photometric H$\alpha$ Survey of the Northern 
Galactic Plane \citep[IPHAS;][]{dre05}.  
\vspace{.1in}\\
{\it HCT and LOT.} 
Follow-up spectroscopic observations were taken using the HFOSC on the 
Himalayan Chandra Telescope (HCT) with a slit width of $1\farcs15$ and Grism 8 
($\lambda = 5800$--$8350$~\AA, dispersion = 1.45~\AA~pixel$^{-1}$) on 
2012 July 26.  Additional spectra were taken using the Hiyoyu spectrograph on 
the Lulin One-meter Telescope (LOT) with a slit width of $1\farcs5$ and grating 
of 300~mm$^{-1}$ ($\lambda = 3800$--$7600$~\AA) on the night of 2012 November 
15.  A one-dimensional spectrum was extracted from the dark, bias-subtracted 
and flat-field-corrected image in the standard manner using IRAF.  The 
wavelength calibration was done using FeNe (HCT) or HeNeAr (LOT) lamp 
sources.

\section{Characterization of the Cluster}
G144 was first recognized by \citet{glu10}, who used the pipeline developed by 
\citet{kop08} to search for density peaks in a star catalog as star clusters.  
They found 153 previously unknown clusters within the Galactic latitude 
$|b|<24\degr$, 23 of which were considered to be embedded star clusters and 
were not characterized by their pipeline.  This candidate was also detected as 
a density peak by our pipeline.  In this section, we present the properties of 
the cluster derived from our analysis.  

\subsection{Density Enhancements and Radial Density Profile}
\begin{figure}
\plotone{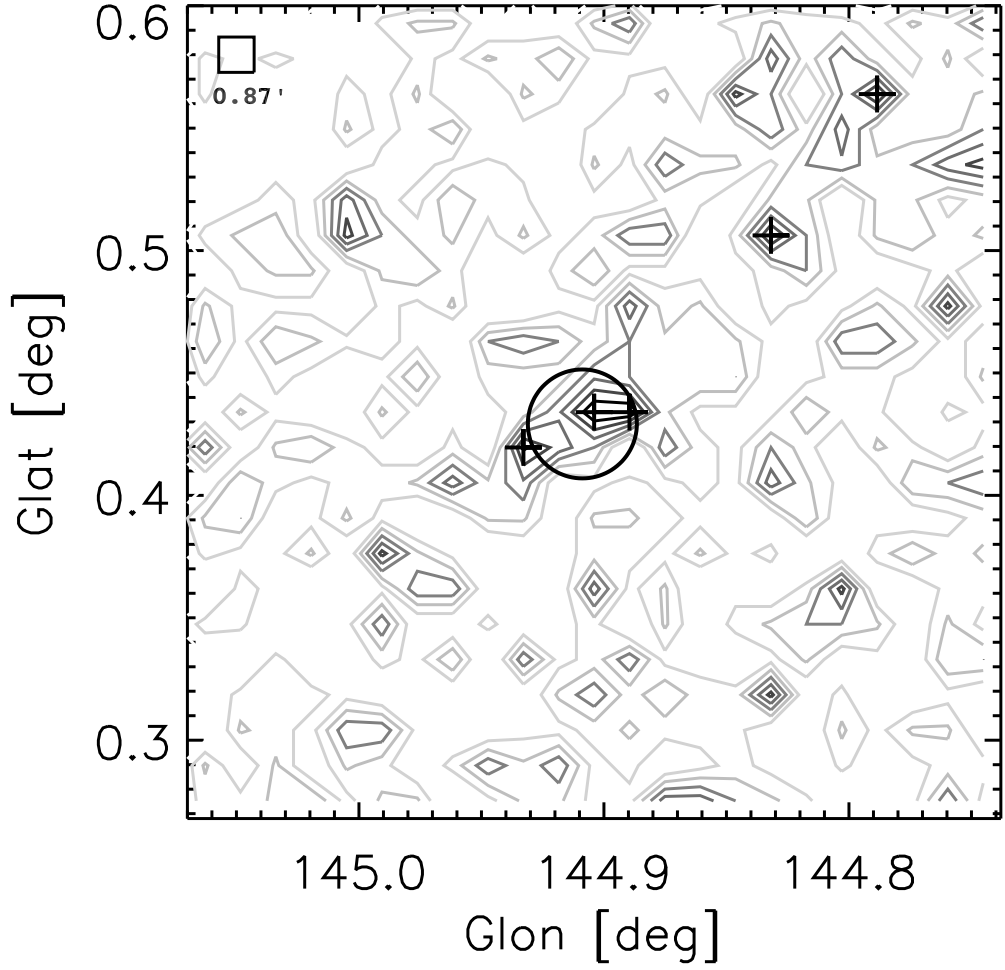}
\caption{ Stellar density map of G144.  The pluses mark the grids 3$\sigma$ 
above the average density.  The circle shows the effective size, which is 
estimated from the three neighboring pluses.  }
\label{dmap}
\end{figure}

We selected the 2MASS point sources with $\mbox{S/N}\ge5$ in $J, H, K$ bands 
within 15\arcmin\ of the apparent center of the cluster.  Figure~\ref{dmap} 
shows the stellar density map for which the number of stars in a grid, whose 
scale was chosen to include an average of 10 stars, was counted.  The resulting 
density map was then smoothed with a $3\times3$ grid width.  For G144, a 
background density of $\sim3.1$~stars~arcmin$^{-2}$ was estimated by $3\sigma$ 
clipping of the density outliers.  A grid in the smoothed density map $3\sigma$ 
above the background would be considered a high-density region, and a 
grouping of more than three connected high-density regions would be identified 
as a density enhancement, i.e., a cluster candidate.  The density map around 
G144 is shown in Figure~\ref{dmap}.  

\begin{figure}
\plotone{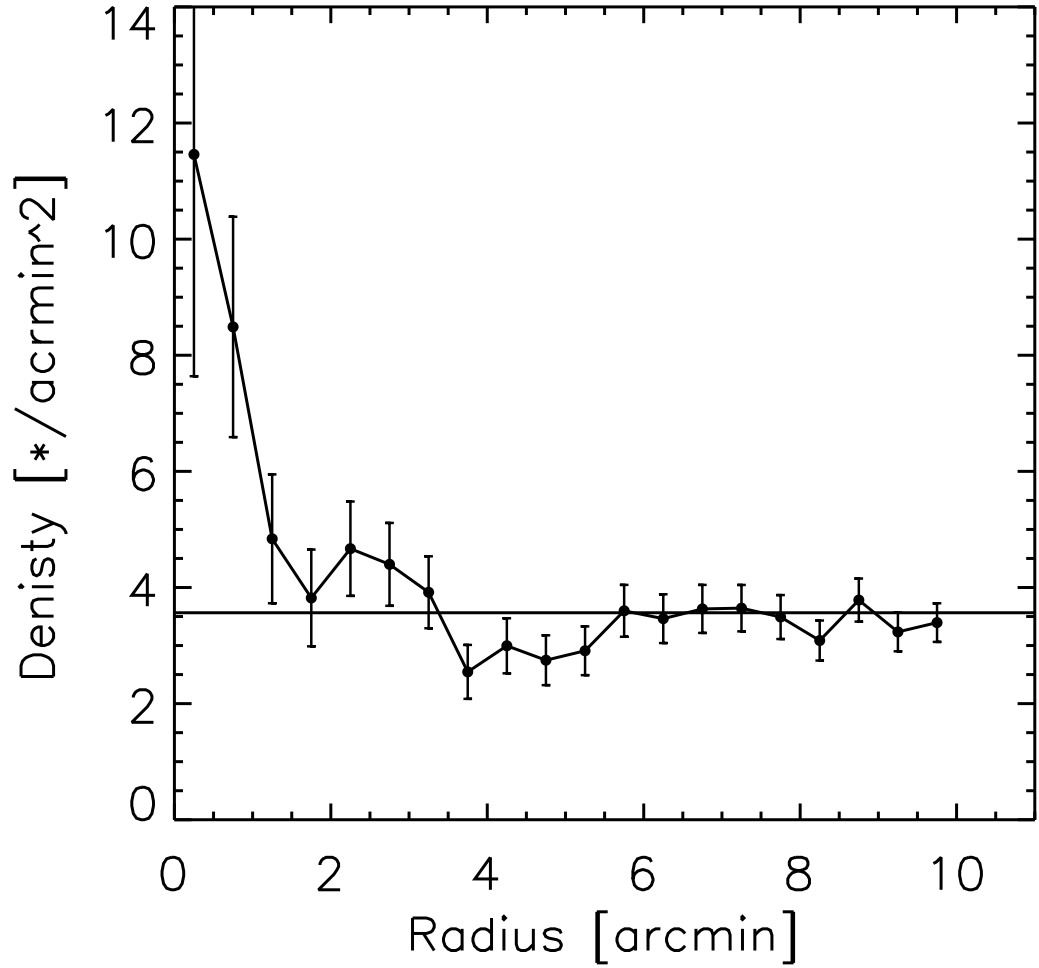}
\caption{ Radial density profile for G144.  Filled dots are surface stellar 
density values with each error bar estimated by Poisson statistics.  The 
horizontal line depicts the background density. The oscillation between the 
radius of 2\arcmin\ and 4\arcmin\ is due to the elongated shape of the 
cluster.  }
\label{rprofile}
\end{figure}

We estimated the size of G144 based on the radial density profile, though the 
cluster appears slightly elongated in shape.  As seen in Figure~\ref{rprofile}, 
the density starts to blend with the field at a radius of $\sim3\farcm5$.  We 
therefore adopted a radius of $3\farcm5$ for the cluster in subsequent 
analysis.  Spatially the cluster is seen near the Cam\,OB1 association and is 
likely physically related to it.  

\subsection{Proper Motion Members}

Cluster members share the same age, metallicity, distance, PMs, etc.  
Accordingly, common PMs would be a necessary condition for cluster membership.  
The Cam\,OB1 association was first investigated by \citet{mor53}, who found 
eight early-type stars aggregated at a distance of $\sim900$~pc.  Several 
studies , e.g., \citet{hau70}, \citet{hum78}, and \citet{lyd01}, append the 
members of Cam\,OB1.  However, the Cam\,OB1 members spread over an area on the 
sky of $\sim20\times10$~deg$^2$ corresponding to a physical dimension of 
$\sim320\times160$~pc$^2$, making a coeval formation of the Cam\,OB1 impossible 
\citep{zee99}.  \citet{str07a} divided the association into three 
subgroups---1A, 1B, and 1C---and found them all at nearly the same distance of 
$\sim1$~kpc. G144 is located between 1A and 1B, which contain 30 and 13 OB 
stars, respectively \citep{str07a}. The other subgroup, 1C, has 76 known 
members mainly from NGC\,1502 \citep{wei97}.    

To provide the criteria for selecting PM members in G144, we first extracted 
the PPMXL PMs of the known OB stars.  The same data would also allow us to 
identify faint OB members that might have escaped recognition in previous 
works. The derived PPMXL PMs of each subgroup are 
$(\mu_{\alpha}$,~$\mu_{\delta})=(-1.1\pm2.1$,~$-1.9\pm1.5)$~mas~yr$^{-1}$ 
for 1A, 
$(\mu_{\alpha}$,~$\mu_{\delta})=(-0.8\pm0.6$,~$-2.3\pm1.0)$~mas~yr$^{-1}$ 
for 1B, and 
$(\mu_{\alpha}$,~$\mu_{\delta})=(-0.2\pm1.5$,~$-0.4\pm2.5)$~mas~yr$^{-1}$ 
for 1C.  Indeed the three subgroups share similar PMs, suggesting a physical 
association.  The parameters for the three subgroups, including the PMs, are 
summarized in Table~\ref{tbl-ppm}.  

\begin{deluxetable}{cccccccr}
\tablewidth{0pt} 
\tabletypesize{\footnotesize}
\tablecaption{Three Subgroups of the Cam\,OB1 Association \label{tbl-ppm}}
\tablehead{
\colhead{Subgroup} & 
\colhead{$\ell$, $b$} & 
\colhead{Angular Size} & 
\colhead{Distance} & 
\colhead{Age} & 
\colhead{No. of Stars} & 
\colhead{$\mu_{\alpha}$\tablenotemark{d}} & 
\colhead{$\mu_{\delta}$\tablenotemark{d}} \\  \vspace{-0.2cm} 
  & 
\colhead{(deg)} & 
\colhead{(deg)} & 
\colhead{(pc)} & 
\colhead{(Myr)} & 
  & 
\colhead{(mas yr$^{-1}$}) & 
\colhead{(mas yr$^{-1}$}) }
\startdata
1A& $140.0,+1.5$& $\sim10\times10$& $1010\pm210$& $\lesssim10$\tablenotemark{a}
& 30\tablenotemark{a}& $-1.1\pm2.1$& $-1.9\pm1.5$\\
1B& $148.0,-0.5$& $\sim10\times10$& $\sim1000$  & $\lesssim10$\tablenotemark{a}
& 13\tablenotemark{a}& $-0.8\pm0.6$& $-2.3\pm1.0$\\
1C& $143.7,+7.7$& 0.1             & $1180\pm160$& $\sim10$\tablenotemark{c}    
& 76\tablenotemark{b}& $-0.2\pm1.5$& $ 0.4\pm2.5$\\
\enddata
\tablenotetext{a}{\citet{str07a}.}
\tablenotetext{b}{Mostly NGC~1502 members \citep{wei97}.}
\tablenotetext{c}{\citet{pau05}.}
\tablenotetext{d}{Derived from PPMXL.}
\end{deluxetable}

\begin{figure}
\plotone{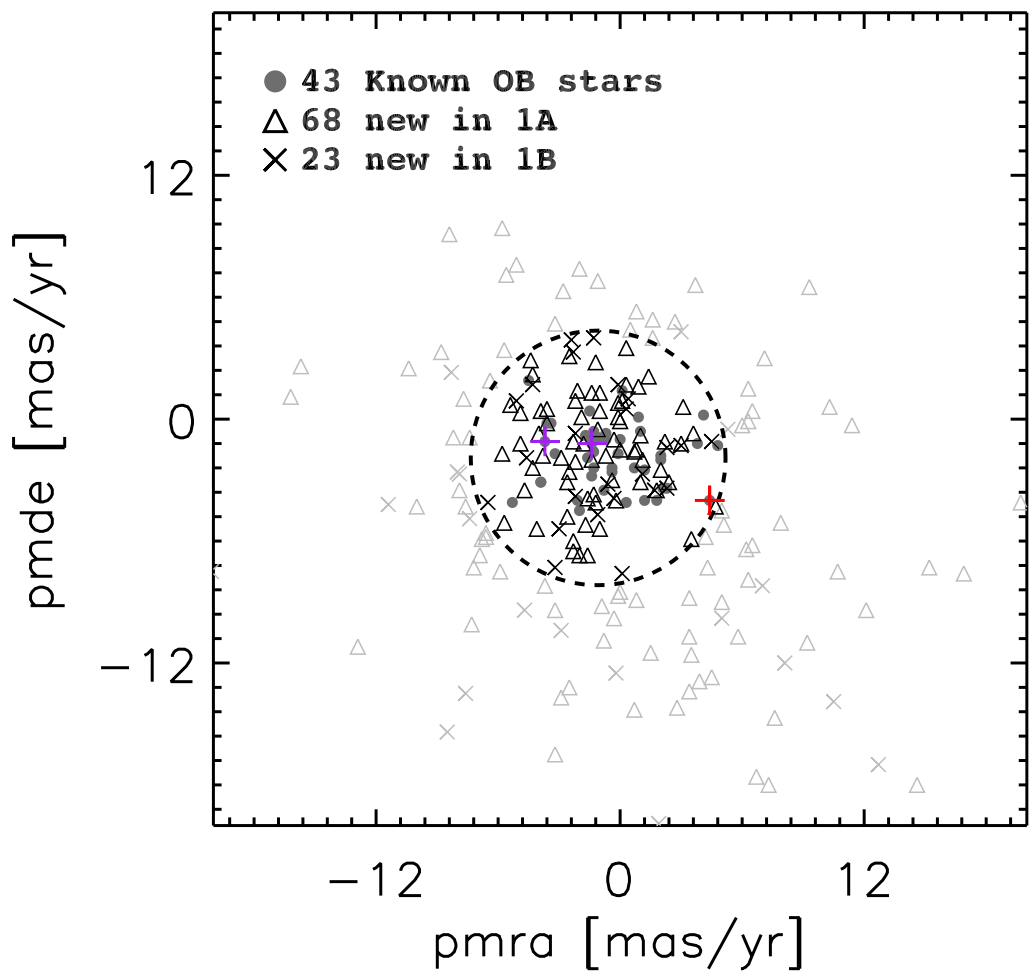}
\caption{ Newly identified OB stars in Cam\,OB1A and 1B.  The filled circles 
mark known Cam\,OB1 members \citep{str07a}, and the two purple and one red 
pluses are supergiants with $J-H$ larger than 0.5~mag (discussed in 
Session~\ref{age}).  The dashed circle is the 3$\sigma$ range of the mean PMs 
of known OB stars in Cam\,OB1A and 1B.  Triangles are OB star candidates in 1A 
and crosses are those in 1B.  Symbols in black represent those with PMs 
consistent with being members.  }
\label{ppmob}
\end{figure}

To identify previously unknown OB stars, we select stars within the same area 
$\sim10\times10$~deg$^2$, as that of the known Cam\,OB1 association centered on 
Cam\,OB1A and 1B, respectively, which are (1)~photometrically consistent with 
OB spectral types in the NIR color--magnitude diagram (CMD), and 
(2)~kinematically consistent with the systemic PM of the 1A and 1B subgroups.  
The CMD criteria are 2MASS $J$ brighter than 13.0~mag and $J-H < 0.2$~mag.  
Only stars with photometric uncertainties better than 0.02~mag in every 2MASS 
band were considered.  The PM criteria include uncertainties less than 
5~mas~yr$^{-1}$ and within the 3$\sigma$ range of the mean PM of 1A and 1B, as 
shown in Figure~\ref{ppmob}.  Table~\ref{tbl-ob} lists the 68 and 23 newly 
found OB stars in 1A and 1B, respectively.  The first column gives the 
numerical identification of the stars, in order of descending brightness, 
followed bysubsequent columns listing the Galactic coordinates, 2MASS 
magnitudes, and PPMXL PMs of each star.  The last column lists the data 
belonging in 1A or 1B.  The photometric and color properties of these newly 
found OB stars will be discussed later together with the cluster members.

\begin{figure}
\plotone{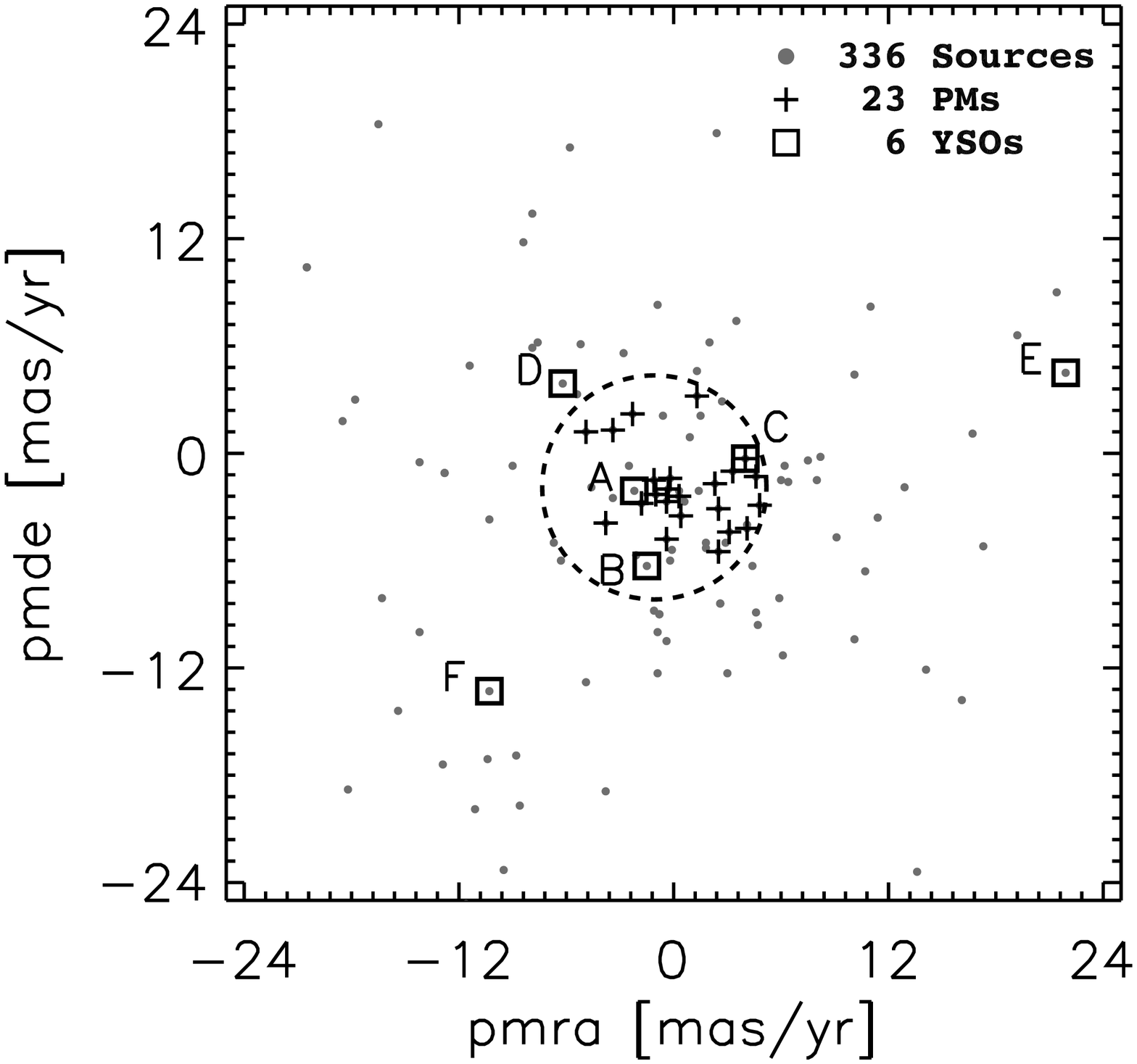}
\caption{ Stellar proper motions toward G144.  Stars within a 3\farcm5 radius 
of the cluster are represented by gray dots.  The dashed circle, as in 
Figure~\ref{ppmob}, shows the probable PM range for membership.  Pluses mark PM 
candidates.  Boxes, individually labeled, are stars with infrared excess or 
highly reddened colors (discussed in Session~\ref{yso}).  }
\label{ppmg144}
\end{figure}

For cluster members, there are 23 stars that are spatially within the 3\farcm5 
radius and kinematically within the 3$\sigma$ range of the average PM of the 
known OB stars, as shown in Figure~\ref{ppmg144} and listed in 
Table~\ref{tbl-ppmem}.  Again, only stars with PPMXL uncertainties less than 
5~mas~yr$^{-1}$ were selected.  The PM candidates need to be further winnowed 
by photometric selection as discussed in the next section.  

\begin{deluxetable}{ccrrrrrc}
\tablecolumns{7} 
\tablewidth{0pt} 
\tabletypesize{\scriptsize}
\tablecaption{New OB Candidates in Cam\,OB 1A and 1B\label{tbl-ob}}
\tablehead{
\colhead{ID} & 
\colhead{$\ell, b$} & 
\colhead{$J$} & 
\colhead{$H$} & 
\colhead{$K$} & 
\colhead{$\mu_\alpha$} & 
\colhead{$\mu_\delta$} & 
\colhead{Comments} \\  \vspace{-0.2cm} 
 & 
\colhead{(deg)} & 
\colhead{(mag)} & 
\colhead{(mag)} & 
\colhead{(mag)} & 
\colhead{(mas~yr$^{-1}$}) & 
\colhead{(mas~yr$^{-1}$}) }
\startdata
  1&$136.423,-0.948$&  6.68&  6.53&  6.44&$-0.1\pm 0.9$&$ 0.1\pm 0.7$&1A\\ 
  2&$135.499,+0.833$&  7.31&  7.14&  7.02&$-3.6\pm 1.1$&$-0.2\pm 0.8$&1A\\ 
  3&$139.336,+2.513$&  8.31&  8.31&  8.26&$-5.8\pm 1.0$&$-1.7\pm 0.9$&1A\\ 
  4&$135.552,+2.559$&  8.73&  8.65&  8.57&$-1.0\pm 1.8$&$-5.4\pm 1.8$&1A\\ 
  5&$135.575,+1.579$&  8.89&  8.82&  8.76&$-0.1\pm 1.4$&$ 0.8\pm 1.1$&1A\\ 
  6&$139.909,+0.635$&  9.16&  9.05&  8.95&$ 0.3\pm 1.8$&$ 1.7\pm 1.7$&1A\\ 
  7&$142.940,-0.745$&  9.23&  9.19&  9.12&$-1.3\pm 1.3$&$-3.7\pm 1.3$&1A\\ 
  8&$139.404,+2.266$&  9.26&  9.17&  9.12&$ 0.8\pm 1.6$&$-1.5\pm 1.6$&1A\\ 
  9&$143.240,-1.364$&  9.29&  9.19&  9.10&$-4.9\pm 1.8$&$ 0.3\pm 1.8$&1A\\ 
 10&$144.540,+4.802$&  9.30&  9.29&  9.25&$-4.3\pm 1.1$&$ 2.2\pm 1.0$&1A\\ 
 11&$140.150,+2.163$&  9.39&  9.37&  9.32&$-2.2\pm 1.6$&$ 0.9\pm 1.7$&1A\\ 
 12&$142.818,+2.377$&  9.55&  9.44&  9.37&$-1.1\pm 2.1$&$-0.5\pm 2.1$&1A\\ 
 13&$135.773,+1.449$&  9.73&  9.65&  9.60&$-3.9\pm 2.0$&$ 0.4\pm 2.0$&1A\\ 
 14&$141.764,-2.567$&  9.79&  9.63&  9.53&$-4.9\pm 2.9$&$-1.2\pm 2.9$&1A\\ 
 15&$139.879,+4.601$&  9.83&  9.75&  9.69&$-1.4\pm 1.9$&$ 1.3\pm 1.9$&1A\\ 
 16&$144.565,+1.886$&  9.95&  9.80&  9.71&$-4.4\pm 1.6$&$ 2.9\pm 1.6$&1A\\ 
 17&$140.789,+2.781$&  9.95&  9.78&  9.68&$ 0.3\pm 2.0$&$ 3.5\pm 2.0$&1A\\ 
 18&$143.333,-0.103$&  9.96&  9.82&  9.76&$-0.2\pm 2.1$&$-4.0\pm 2.1$&1A\\ 
 19&$144.446,-1.431$& 10.04&  9.86&  9.76&$-2.9\pm 2.1$&$-1.9\pm 2.1$&1A\\ 
 20&$143.361,+3.974$& 10.06&  9.95&  9.87&$-0.3\pm 1.7$&$-1.0\pm 1.7$&1A\\ 
 21&$143.964,+3.254$& 10.25& 10.17& 10.12&$ 1.1\pm 2.1$&$-2.0\pm 2.1$&1A\\ 
 22&$144.320,+5.428$& 10.27& 10.28& 10.28&$-4.3\pm 1.6$&$-2.4\pm 1.6$&1A\\ 
 23&$137.583,+4.872$& 10.35& 10.15& 10.10&$ 1.8\pm 2.0$&$-3.5\pm 2.0$&1A\\ 
 24&$143.845,+5.805$& 10.37& 10.22& 10.17&$-1.2\pm 1.9$&$ 2.8\pm 1.9$&1A\\ 
 25&$136.891,+5.032$& 10.40& 10.20& 10.15&$-1.0\pm 1.7$&$ 1.3\pm 1.7$&1A\\ 
 26&$139.130,+1.675$& 10.41& 10.22& 10.14&$-2.5\pm 1.7$&$-2.6\pm 1.7$&1A\\ 
 27&$141.658,+1.215$& 10.46& 10.35& 10.27&$-5.7\pm 2.3$&$-5.1\pm 2.3$&1A\\ 
 28&$143.742,+1.059$& 10.46& 10.32& 10.24&$-1.0\pm 2.2$&$ 0.4\pm 2.2$&1A\\ 
 29&$143.141,+3.525$& 10.46& 10.38& 10.29&$-1.9\pm 1.8$&$ 0.1\pm 1.8$&1A\\ 
 30&$138.041,-0.551$& 10.49& 10.40& 10.31&$-1.8\pm 2.1$&$-1.2\pm 2.1$&1A\\ 
 31&$143.335,-0.101$& 10.50& 10.37& 10.27&$-0.4\pm 2.2$&$-3.0\pm 2.2$&1A\\ 
 32&$144.343,+4.769$& 10.51& 10.35& 10.28&$-2.5\pm 1.7$&$ 3.1\pm 1.7$&1A\\ 
 33&$143.773,+2.906$& 10.56& 10.42& 10.31&$ 0.0\pm 1.8$&$-1.5\pm 1.8$&1A\\ 
 34&$143.795,+3.308$& 10.56& 10.42& 10.35&$ 0.7\pm 1.8$&$-1.6\pm 1.8$&1A\\ 
 35&$139.743,+4.774$& 10.59& 10.44& 10.29&$ 3.6\pm 2.7$&$-0.7\pm 2.7$&1A\\ 
 36&$140.181,+2.158$& 10.63& 10.52& 10.48&$-4.1\pm 2.0$&$-5.4\pm 2.0$&1A\\ 
 37&$141.804,+2.273$& 10.66& 10.55& 10.48&$ 3.5\pm 1.8$&$-5.9\pm 1.8$&1A\\ 
 38&$144.248,+2.220$& 10.67& 10.54& 10.47&$ 0.9\pm 1.8$&$ 1.6\pm 1.8$&1A\\ 
 39&$142.569,+0.978$& 10.70& 10.53& 10.43&$-1.4\pm 1.9$&$-2.0\pm 1.9$&1A\\ 
 40&$143.314,-0.222$& 10.76& 10.66& 10.60&$ 2.0\pm 1.6$&$-2.5\pm 1.6$&1A\\ 
 41&$138.486,+3.456$& 10.77& 10.59& 10.54&$ 2.4\pm 2.7$&$-3.1\pm 2.7$&1A\\ 
 42&$135.654,+1.584$& 10.78& 10.77& 10.71&$ 1.0\pm 2.0$&$-0.8\pm 2.0$&1A\\ 
 43&$144.849,-2.906$& 10.82& 10.67& 10.48&$-2.2\pm 4.1$&$-2.1\pm 4.1$&1A\\ 
 44&$138.594,+2.357$& 10.85& 10.66& 10.53&$ 2.2\pm 2.8$&$-1.1\pm 2.8$&1A\\ 
 45&$144.165,+3.279$& 10.86& 10.74& 10.67&$-2.1\pm 2.1$&$ 1.4\pm 2.1$&1A\\ 
 46&$137.987,-0.541$& 10.88& 10.69& 10.58&$-1.2\pm 1.8$&$-4.1\pm 1.8$&1A\\ 
 47&$141.849,+2.190$& 10.95& 10.80& 10.73&$ 3.0\pm 2.2$&$-1.2\pm 2.2$&1A\\ 
 48&$143.304,-0.146$& 10.98& 10.90& 10.78&$-2.6\pm 1.8$&$-4.8\pm 1.8$&1A\\ 
 49&$138.642,+2.290$& 10.99& 10.89& 10.82&$-3.6\pm 2.0$&$ 0.5\pm 2.0$&1A\\ 
 50&$144.131,-2.173$& 11.01& 10.89& 10.82&$ 0.1\pm 1.8$&$ 0.9\pm 1.8$&1A\\ 
 51&$143.997,+3.374$& 11.02& 10.84& 10.74&$-1.6\pm 2.1$&$-3.9\pm 2.1$&1A\\ 
 52&$141.297,+2.745$& 11.04& 10.91& 10.83&$-2.6\pm 2.1$&$-3.1\pm 2.1$&1A\\ 
 53&$143.963,-1.950$& 11.07& 10.93& 10.87&$-4.7\pm 2.3$&$-3.5\pm 2.3$&1A\\ 
 54&$144.713,+2.623$& 11.09& 10.90& 10.80&$-2.3\pm 2.2$&$-1.1\pm 2.2$&1A\\ 
 55&$144.094,+2.151$& 11.10& 11.01& 10.91&$-5.4\pm 1.8$&$ 0.7\pm 1.9$&1A\\ 
 56&$143.235,+4.009$& 11.16& 11.04& 10.95&$-2.3\pm 1.7$&$-6.0\pm 1.7$&1A\\ 
 57&$143.371,-0.256$& 11.19& 11.07& 11.02&$ 4.7\pm 1.7$&$-4.3\pm 1.7$&1A\\ 
 58&$143.492,-1.700$& 11.22& 11.06& 10.96&$-2.3\pm 2.3$&$-6.5\pm 2.3$&1A\\ 
 59&$141.780,-2.387$& 11.24& 11.04& 10.95&$ 0.0\pm 2.9$&$-0.1\pm 2.9$&1A\\ 
 60&$137.154,+5.389$& 11.35& 11.16& 11.09&$-1.6\pm 4.1$&$-6.7\pm 4.1$&1A\\ 
 61&$137.392,+3.852$& 11.38& 11.24& 11.16&$ 1.0\pm 1.7$&$-3.1\pm 1.6$&1A\\ 
 62&$144.385,+2.014$& 11.42& 11.24& 11.13&$ 3.1\pm 2.9$&$ 0.6\pm 2.9$&1A\\ 
 63&$143.751,+3.354$& 11.45& 11.26& 11.19&$-4.0\pm 1.8$&$-0.7\pm 1.8$&1A\\ 
 64&$143.183,-1.119$& 11.49& 11.32& 11.21&$ 1.4\pm 2.2$&$ 2.1\pm 2.2$&1A\\ 
 65&$142.926,-0.749$& 11.53& 11.43& 11.33&$-2.0\pm 2.2$&$-6.7\pm 2.2$&1A\\ 
 66&$144.419,+3.040$& 11.53& 11.39& 11.34&$-3.8\pm 1.7$&$-1.8\pm 1.7$&1A\\ 
 67&$138.831,+6.303$& 11.59& 11.46& 11.37&$-0.7\pm 4.0$&$-1.8\pm 4.0$&1A\\ 
 68&$135.975,+1.828$& 11.88& 11.73& 11.63&$-1.7\pm 2.7$&$-5.2\pm 2.7$&1A\\ 
  1&$146.869,-5.801$&  6.04&  5.89&  5.83&$ 0.4\pm 0.6$&$ 1.0\pm 0.7$&1B\\ 
  2&$145.762,+3.327$&  9.32&  9.20&  9.10&$-0.6\pm 1.8$&$-3.2\pm 1.8$&1B\\ 
  3&$146.307,-5.275$&  9.50&  9.43&  9.36&$-0.1\pm 1.4$&$ 1.7\pm 1.4$&1B\\ 
  4&$146.855,-0.543$&  9.54&  9.39&  9.31&$-4.6\pm 2.0$&$-1.9\pm 2.0$&1B\\ 
  5&$146.016,+3.242$&  9.91&  9.85&  9.81&$-2.4\pm 2.1$&$ 3.9\pm 2.1$&1B\\ 
  6&$146.804,-0.548$&  9.94&  9.83&  9.73&$-4.3\pm 2.7$&$ 1.7\pm 2.7$&1B\\ 
  7&$145.503,-6.755$& 10.21& 10.15& 10.07&$ 0.3\pm 1.7$&$ 0.5\pm 1.7$&1B\\ 
  8&$146.309,-0.178$& 10.40& 10.29& 10.19&$-1.3\pm 1.7$&$ 4.0\pm 1.7$&1B\\ 
  9&$151.415,+3.612$& 10.42& 10.31& 10.22&$-2.2\pm 2.0$&$-0.7\pm 2.0$&1B\\ 
 10&$150.306,-6.250$& 10.44& 10.37& 10.28&$ 0.1\pm 1.4$&$-7.6\pm 1.4$&1B\\ 
 11&$151.443,+3.504$& 10.84& 10.64& 10.56&$-6.5\pm 4.0$&$-4.1\pm 4.0$&1B\\ 
 12&$146.497,+3.146$& 10.86& 10.68& 10.60&$ 2.3\pm 1.8$&$-1.4\pm 1.8$&1B\\ 
 13&$148.254,+2.582$& 10.87& 10.75& 10.70&$-2.3\pm 1.8$&$ 3.3\pm 1.8$&1B\\ 
 14&$148.387,+0.963$& 11.07& 10.93& 10.85&$ 4.5\pm 1.7$&$-1.1\pm 1.7$&1B\\ 
 15&$146.421,-5.025$& 11.14& 10.97& 10.84&$-2.2\pm 4.1$&$-3.8\pm 4.1$&1B\\ 
 16&$146.069,+3.633$& 11.52& 11.47& 11.39&$-0.3\pm 1.8$&$-3.9\pm 1.8$&1B\\ 
 17&$148.234,+2.526$& 11.52& 11.39& 11.30&$-5.1\pm 2.2$&$ 0.9\pm 2.2$&1B\\ 
 18&$146.432,-5.113$& 11.59& 11.39& 11.27&$-3.0\pm 4.1$&$-5.4\pm 4.1$&1B\\ 
 19&$151.412,+3.551$& 11.59& 11.43& 11.34&$ 1.1\pm 4.0$&$-2.7\pm 4.0$&1B\\ 
 20&$145.901,-4.777$& 11.72& 11.58& 11.49&$-3.2\pm 4.1$&$-7.3\pm 4.1$&1B\\ 
 21&$145.648,-4.026$& 11.86& 11.67& 11.56&$ 3.0\pm 4.1$&$-1.3\pm 4.1$&1B\\ 
 22&$156.362,-6.956$& 11.99& 11.89& 11.82&$-1.1\pm 3.9$&$-4.7\pm 3.9$&1B\\ 
 23&$145.908,-4.784$& 12.05& 11.87& 11.80&$ 1.7\pm 4.1$&$-3.5\pm 4.1$&1B\\ 
\enddata
\end{deluxetable}

\subsection{Young Stellar Objects \label{yso}}

\citet{str07b} have used infrared photometric data extracted from 2MASS, 
{\it IRAS} and {\it MSX} to identify 35 YSOs in a region $12\times12$~deg$^2$ 
toward the Cam\,OB1 association, none of which are located within G144.  
Nonetheless, nebulosity is clearly seen around G144, suggesting a young age.  
Generally, YSOs are identified by their infrared excess, H$\alpha$ emission, 
or X-ray emission.  No X-ray data are available in the region, so we used 
2MASS and {\it WISE} to identify possible YSOs or embedded stars in G144.

\begin{figure}
\plotone{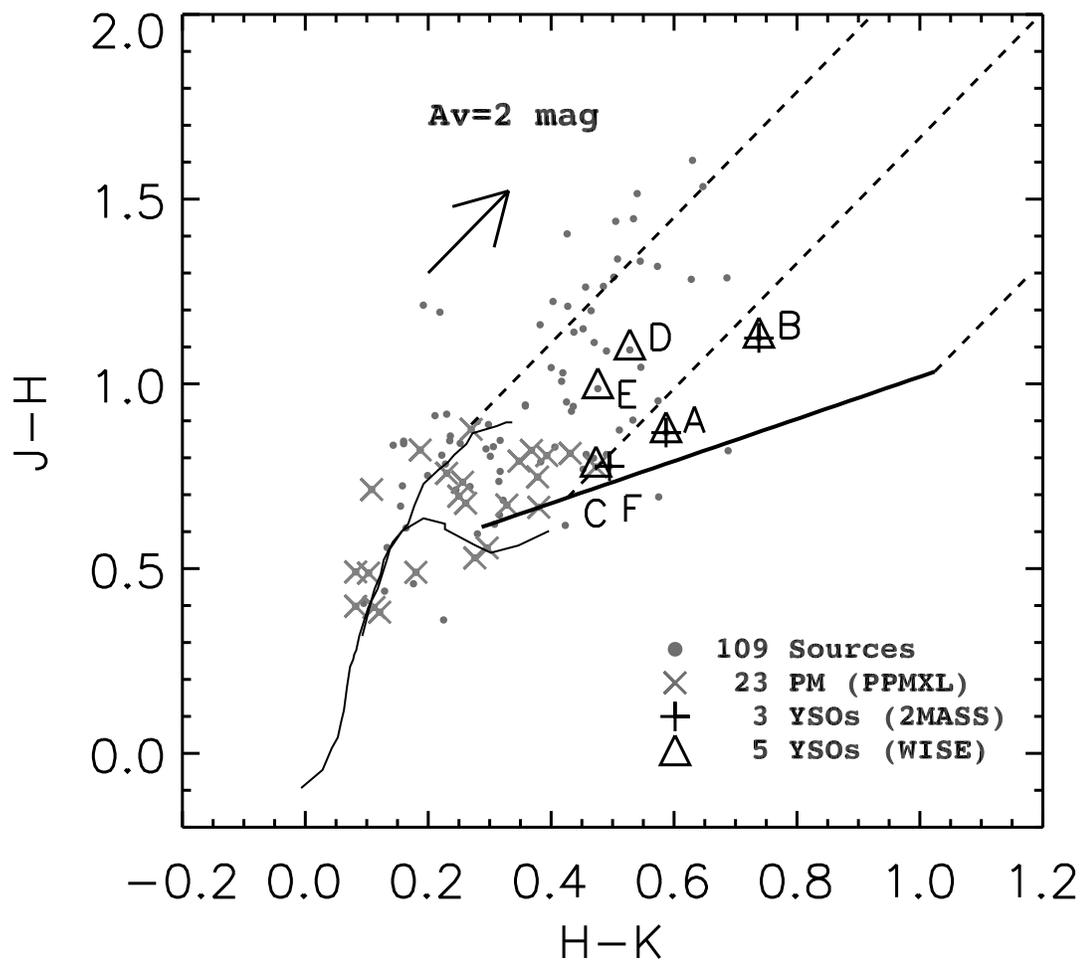}
\caption{ 2MASS color--color diagram in the cluster region.  Pluses and 
triangles are CTTS candidates (pluses denote 2MASS sources and triangles 
denote {\it WISE}).  Crosses represent proper motion members.  Gray dots are 
none of the above.  The thin curves show the giant and dwarf loci 
\citep{bnb88} converted to the 2MASS system.  The thick line is the intrinsic 
CTTS locus \citep{mey97}. The arrow represents the reddening direction 
\citep{rei85} for typical Galactic interstellar extinction ($R_V=3.1$), and 
the dashed lines encompass the region of reddened giants and dwarfs.  Stars to 
the right of this region have NIR excess and are possible YSOs. }  
\label{tmccd}
\end{figure}

\begin{figure}
\plotone{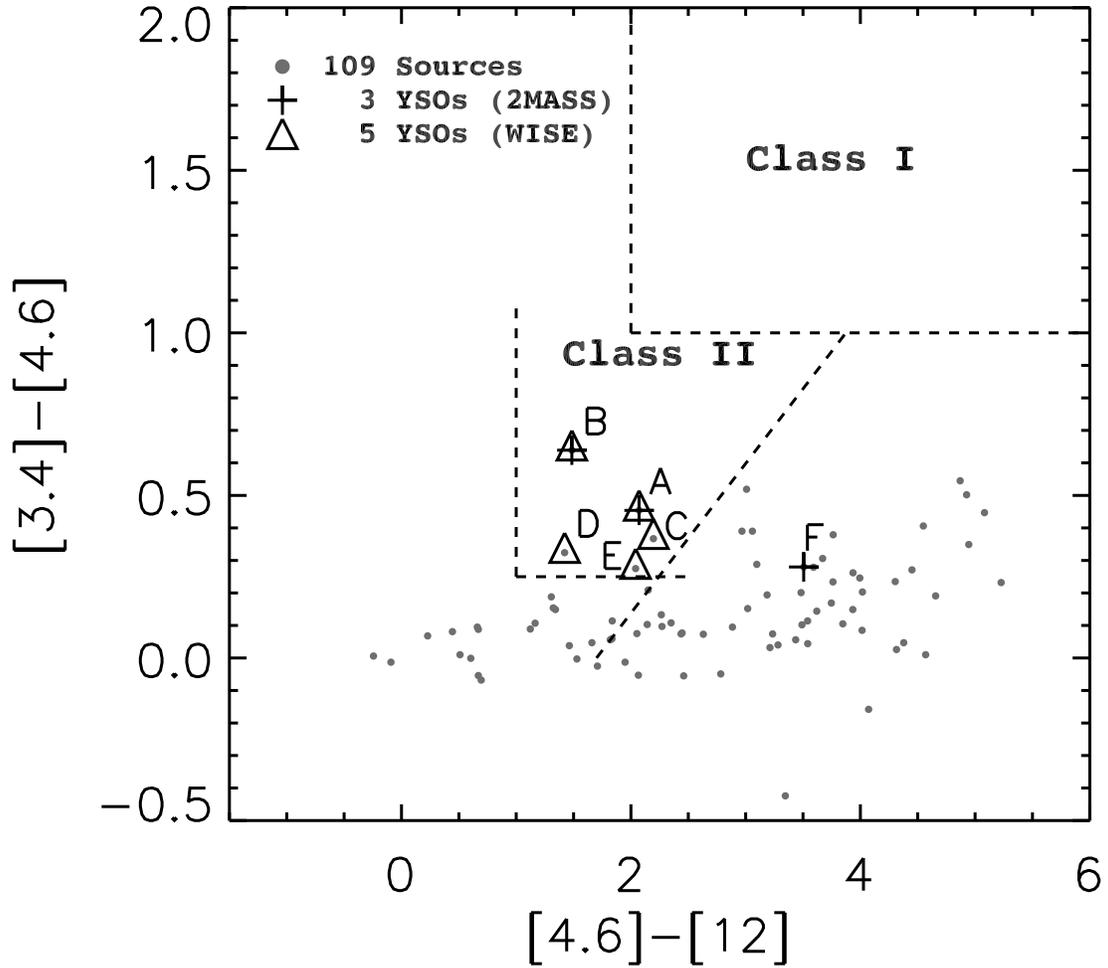}
\caption{ {\it WISE} $[3.4]-[4.6]~\micron$ vs. $[4.6]-[12]~\micron$ colors in 
the cluster region.  The dashed lines defined by \citet{koe12} are used to 
identify Class~I and Class~II young stellar objects.  As in Figure~\ref{tmccd}, 
the 2MASS detections are in pluses and the {\it WISE} detections are in 
triangles.  }
\label{wsccd}
\end{figure}

The NIR-excess sources were identified using the 2MASS $(J-H)$ versus $(H-K)$ 
and {\it WISE} $[3.4]-[4.6]$ versus $[4.6]-[12]$ color-color diagrams, shown 
in Figure~\ref{tmccd} and Figure~\ref{wsccd}.  Three classical T Tauri (CTTS; 
class II objects) candidates were found in the empirical CTTS region 
\citep{mey97} in the 2MASS color-color diagram.  On the other hand, using the 
criteria adopted by \citet{koe12}, five CTTS candidates were identified on 
the basis of the {\it WISE} colors, two of which coincide with the 2MASS 
sources.  Therefore, a total of six CTTS candidates are selected, whose 
coordinates, 2MASS and {\it WISE} magnitudes, and PPMXL PMs are listed in 
Table~\ref{tbl-yso}.  The first column gives the alphabetical identification 
of the star, in order of descending brightness, followed by subsequent columns 
of Galactic coordinates, 2MASS as well as {\it WISE} magnitudes, and PPMXL PM 
of each star.  The last column lists the data used for the classification.   

\begin{deluxetable}{ccrrrrr}
\tablecolumns{7} 
\tablewidth{0pt} 
\tabletypesize{\scriptsize}
\tablecaption{Proper Motion Member Candidates\label{tbl-ppmem}}
\tablehead{
\colhead{ID} & 
\colhead{$\ell, b$} & 
\colhead{$J$} & 
\colhead{$H$} & 
\colhead{$K$} & 
\colhead{$\mu_\alpha$} & 
\colhead{$\mu_\delta$} \\  \vspace{-0.2cm} 
 & 
\colhead{(deg)} & 
\colhead{(mag)} & 
\colhead{(mag)} & 
\colhead{(mag)} & 
\colhead{(mas~yr$^{-1}$}) & 
\colhead{(mas~yr$^{-1}$}) }
\startdata
  1&$144.911,+0.395$& 11.89& 11.50& 11.41&$-0.4\pm 4.0$&$-4.8\pm 4.0$\\
  2&$144.898,+0.437$& 11.93& 11.25& 10.99&$-4.9\pm 4.0$&$ 1.2\pm 4.0$\\
  3&$144.946,+0.409$& 12.28& 11.53& 11.15&$-1.0\pm 4.0$&$-2.3\pm 4.0$\\
  4&$144.913,+0.418$& 12.71& 12.04& 11.71&$-0.3\pm 4.0$&$-2.0\pm 4.0$\\
  5&$144.903,+0.437$& 12.96& 12.15& 11.76&$ 1.3\pm 4.0$&$ 3.2\pm 4.0$\\
  6&$144.891,+0.391$& 13.12& 12.74& 12.62&$ 0.3\pm 4.0$&$-2.4\pm 4.0$\\
  7&$144.918,+0.420$& 13.14& 12.35& 12.00&$ 4.1\pm 4.0$&$-4.2\pm 4.0$\\
  8&$144.909,+0.462$& 13.40& 12.91& 12.81&$ 2.5\pm 4.0$&$-5.5\pm 4.0$\\
  9&$144.940,+0.402$& 13.43& 12.94& 12.86&$-3.4\pm 4.0$&$ 1.3\pm 4.0$\\
 10&$144.858,+0.449$& 13.66& 13.26& 13.15&$-0.2\pm 4.0$&$-1.4\pm 4.0$\\
 11&$144.917,+0.465$& 13.69& 13.20& 13.02&$ 3.1\pm 4.0$&$-4.4\pm 4.0$\\
 12&$144.896,+0.430$& 13.94& 13.16& 12.69&$ 4.0\pm 4.0$&$-0.3\pm 4.0$\\
 13&$144.913,+0.468$& 14.58& 13.91& 13.53&$-1.8\pm 4.0$&$-2.8\pm 4.0$\\
 14&$144.955,+0.436$& 14.63& 13.87& 13.64&$-3.8\pm 4.0$&$-3.9\pm 4.0$\\
 15&$144.928,+0.430$& 14.67& 13.97& 13.72&$-1.1\pm 4.0$&$-1.5\pm 4.0$\\
 16&$144.942,+0.452$& 14.73& 13.92& 13.49&$ 3.3\pm 4.0$&$-1.0\pm 4.0$\\
 17&$144.862,+0.464$& 14.85& 14.29& 14.00&$ 4.8\pm 4.0$&$-2.9\pm 4.0$\\
 18&$144.877,+0.401$& 14.90& 14.19& 14.08&$-0.4\pm 4.0$&$-2.7\pm 4.0$\\
 19&$144.906,+0.455$& 14.92& 14.10& 13.91&$ 2.5\pm 4.0$&$-3.1\pm 4.0$\\
 20&$144.872,+0.473$& 14.92& 14.39& 14.11&$ 4.6\pm 4.0$&$-1.3\pm 4.0$\\
 21&$144.948,+0.453$& 15.13& 14.31& 13.94&$ 0.4\pm 4.0$&$-3.5\pm 4.0$\\
 22&$144.915,+0.456$& 15.43& 14.69& 14.44&$ 2.3\pm 4.1$&$-1.7\pm 4.1$\\
 23&$144.866,+0.404$& 15.91& 15.04& 14.77&$-2.3\pm 4.1$&$ 2.2\pm 4.1$\\
 \enddata
\end{deluxetable}

\begin{deluxetable}{ccrrrrrrcrrl}
\tabletypesize{\scriptsize}
\tablecolumns{7} 
\tablewidth{0pt} 
\tablecaption{YSO Candidates in G144 \label{tbl-yso}}
\tablehead{
\colhead{ID} & 
\colhead{$\ell, b$} & 
\colhead{$J$} & 
\colhead{$H$} & 
\colhead{$K$} & 
\colhead{$W1$} & 
\colhead{$W2$} & 
\colhead{$W3$} & 
\colhead{$W4$} & 
\colhead{$\mu_\alpha$} & 
\colhead{$\mu_\delta$} & 
\colhead{Comments} \\ \vspace{-0.2cm}
 & 
\colhead{(deg)} & 
\colhead{(mag)} & 
\colhead{(mag)} & 
\colhead{(mag)} & 
\colhead{(mag)} & 
\colhead{(mag)} & 
\colhead{(mag)} & 
\colhead{(mag)} & 
\colhead{(mas~yr$^{-1}$}) & 
\colhead{(mas~yr$^{-1}$})}
\startdata
A& $144.895,+0.439$ & 11.4& 10.6& 10.0&  9.4&  9.0&  6.9& 2.9& $ -2.2\pm5.0$ &
$ -2.1\pm5.0$ & {\it WISE}; 2MASS\\
B& $144.881,+0.403$ & 13.4& 12.3& 11.6& 10.5&  9.8&  8.3& 6.6& $ -1.5\pm5.3$ &
$ -6.3\pm5.3$ & {\it WISE}; 2MASS\\
C& $144.896,+0.430$ & 13.9& 13.2& 12.7& 12.1& 11.7&  9.5& 5.3& $  4.0\pm4.0$ &
$ -0.3\pm4.0$ & {\it WISE}\\
D& $144.899,+0.426$ & 14.2& 13.1& 12.6& 11.9& 11.6& 10.2& 6.1& $ -6.2\pm5.4$ &
$  3.9\pm5.4$ & {\it WISE}\\
E& $144.914,+0.412$ & 16.1& 15.1& 14.6& 14.5& 14.2& 12.1& 8.7& $ 21.9\pm5.5$ &
$  4.5\pm5.5$ & {\it WISE}; non-member\\
F& $144.853,+0.454$ & 16.2& 15.4& 14.9& 14.8& 14.5& 11.0& 8.6& $-10.3\pm5.5$ &
$-13.3\pm5.5$ & 2MASS; non-member\\
\enddata
\end{deluxetable}

\begin{figure}
\plotone{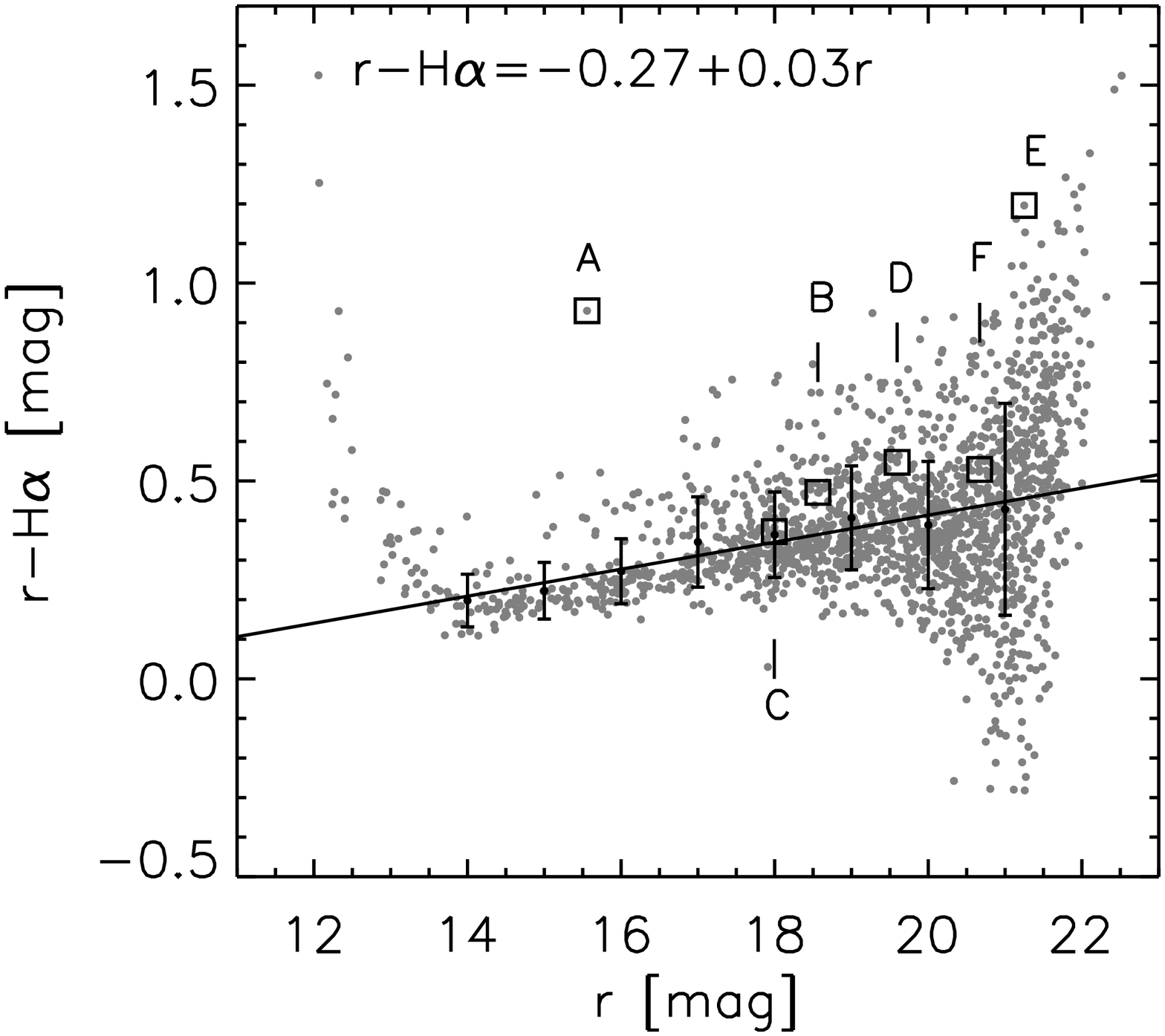}
\caption{ IPHAS magnitudes of $r-$H$\alpha$ vs $r$.  Each YSO candidate is 
labeled and marked with a box.  Bright stars ($r<$ 14~mag) are saturated.  The 
typical error, marked for each magnitude, has been estimated by Poisson 
statistics.  The line shows the linear regression of $r-$H$\alpha$ and $r$ 
magnitudes of stars between $r\sim$ 14--21~mag.  }
\label{halpha}
\end{figure}
\begin{figure}
\plotone{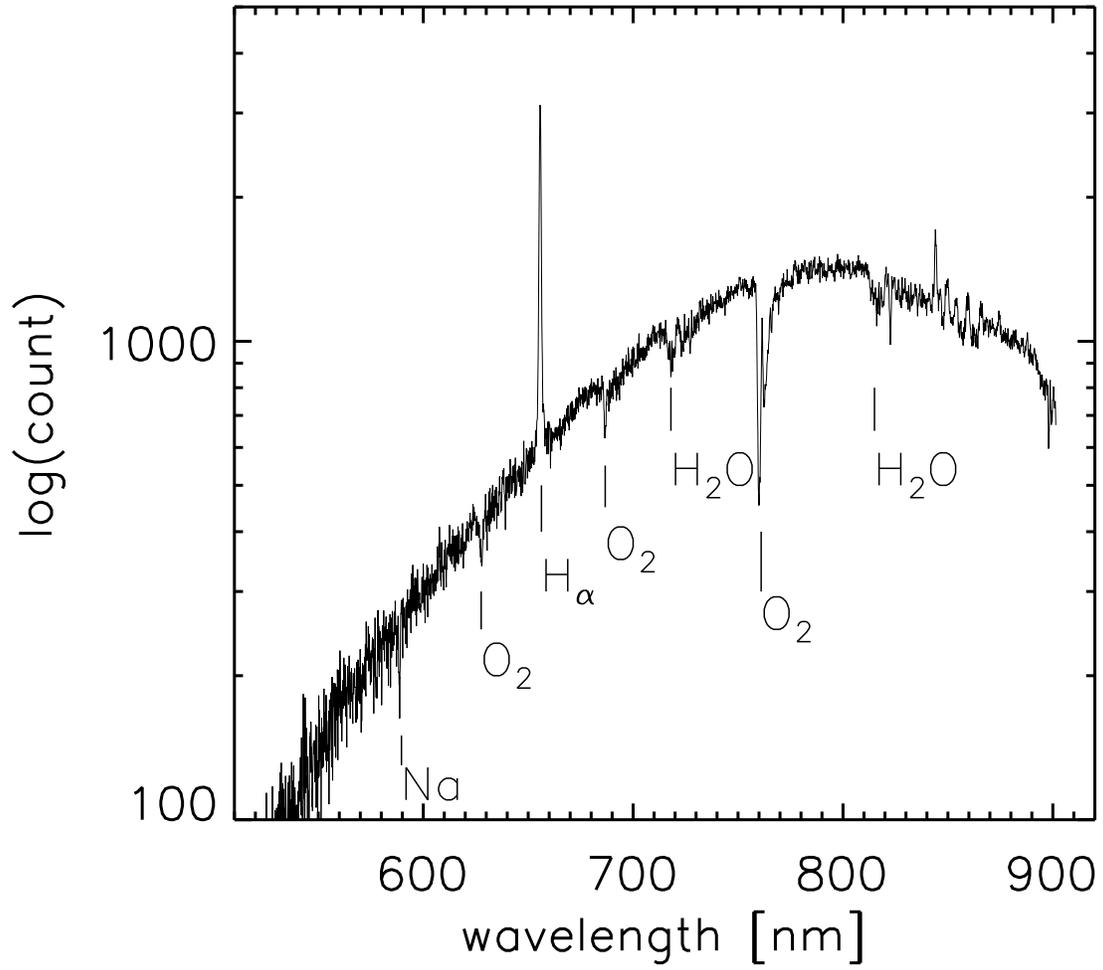}
\caption{ HCT spectrum of star A.  A prominent H$\alpha$ emission line is 
evident.  Other spectral features are also marked.  }
\label{stara}
\end{figure}

The IPHAS $r$ and narrow-band H$\alpha$ magnitudes are used to single out stars 
with possible H$\alpha$ emission.  Figure~\ref{halpha} shows the $r-$H$\alpha$ 
versus $r$-band magnitude.  Star A, the brightest among our CTTS candidates, 
is some $\sim0.5$~mag brighter in H$\alpha$ than a ``normal'' star, defined as 
a star following the trend between the H$\alpha$- and $r$-band magnitudes for 
the majority of stars in the field.  Subsequent HCT spectroscopic observations 
confirmed an emission spectrum for star A, as seen in Figure~\ref{stara}.  On 
the basis of the NIR excess and the emission-line spectrum, we conclude that 
star A, located near the center of G144, should be a CTTS.  

For other YSO candidates, star B, located near the nebulosity edge, has 
the most reddened NIR colors among our candidates, and is a CTTS candidate 
judging by both 2MASS and {\it WISE} colors.  It may have weak H$\alpha$ 
emission (see Figure~\ref{halpha}), but, unfortunately, was too faint for our 
spectroscopic observations.  Star B is likely a CTTS.  Both stars C and D are 
both marginal candidates in terms of their PMs and {\it WISE} colors.  Their 
2MASS colors do not stand out, though star D, with possible H$\alpha$ in 
emission, is much more reddened than star C.  Both stars E and F are very 
faint, so have large uncertainties in their PM measurements.  They are 
marginal cases in either 2MASS or {\it WISE} colors, and seem to have motions 
inconsistent with membership.  Hence, stars E and F should be classified as 
field stars.  

\subsection{Extinction, Distance, and Age \label{age}}

\begin{figure}
\plotone{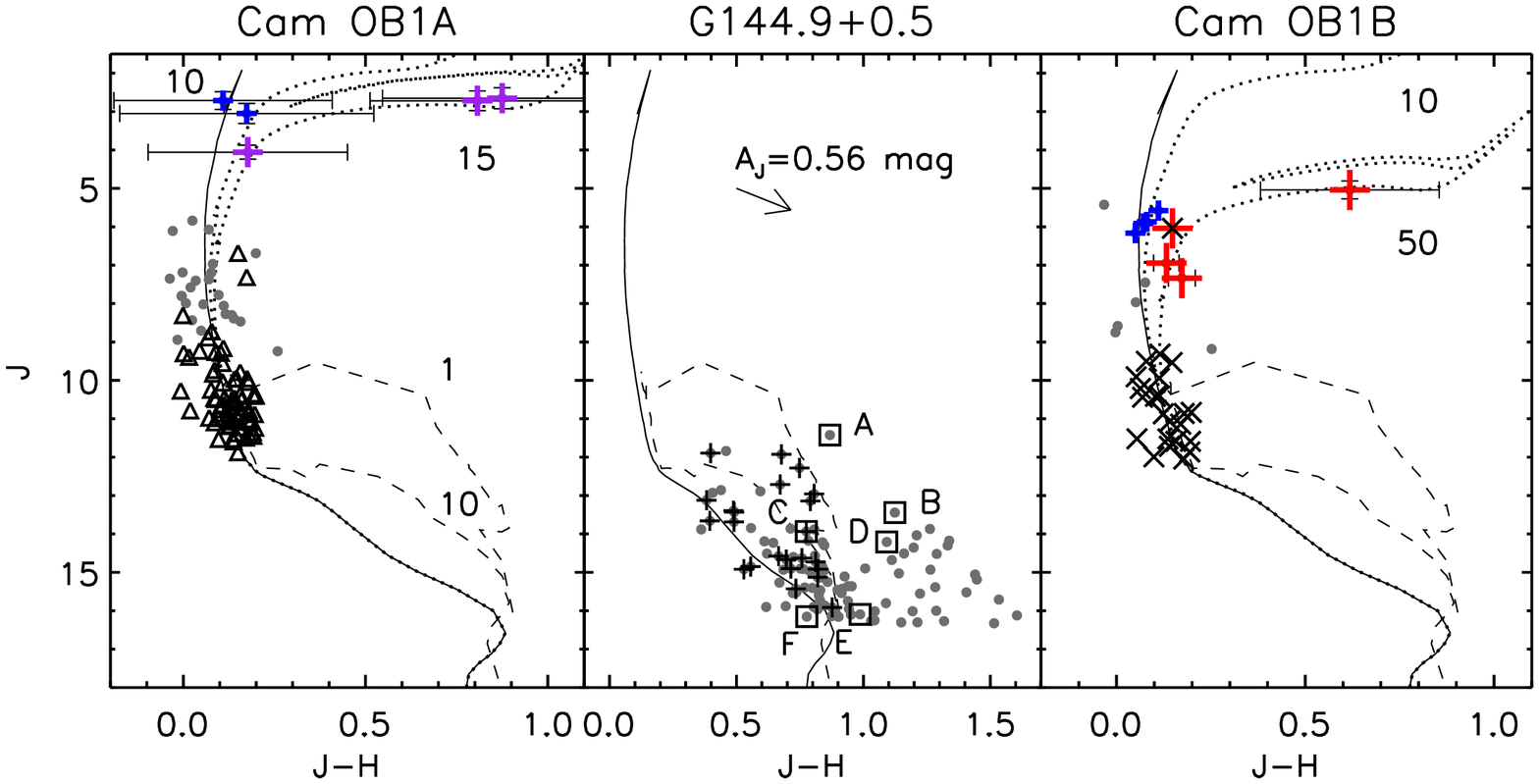}
\caption{ $J$ vs. $J-H$ diagrams for Cam\,OB 1A (left), 1B (right), and 
G144 (middle).  The solid line in each case delineates the ZAMS located at 
1~kpc with an interstellar extinction of $A_V\sim2.0$~mag.  The dashed lines 
represent the pre-main-sequence isochrone for 1 and 10~Myr \citep{sis00}.  
The dotted lines represent the post-main-sequence isochrone for 
10 and 15~Myr (left) and 10 and 50~Myr \citep[rigth;][]{gir02}.  The gray dots 
in the left or right panels mark known 1A and 1B members, whereas those in the 
middle panel are stars in the cluster region.  The newly found OB stars are 
represented by triangles (1A) and crosses (1B); purple, blue, and red pluses 
are supergiants (discussed in Session~\ref{age}).  In the cluster region, 
the pluses show the 23 PM candidates whereas the boxes represent YSO 
candidates (labeled).  }
\label{obcmd}
\end{figure}

The average interstellar extinction to the cluster was estimated by tracing 
each of the 23 PM candidates back to the MS locus along the reddening vector in 
the 2MASS color-color diagram (see Figure~\ref{tmccd}).  An average 
extinction of $A_J\sim 0.56$~mag was derived.  Figure~\ref{obcmd} plots the $J$ 
versus $J-H$ CMD for the Cam\,OB1 subgroups 1A and 1B, as well as for G144.  In 
each case, a zero-age main-sequence (ZAMS) is shown with solar 
metallicity\footnote{Available from the Padova isochrone database 
\url{http://stev.oapd.inaf.it/cgi-bin/cmd}.} at a distance of $\sim1$~kpc with 
corresponding extinction and reddening.  

The 1C subgroup has a well constrained age ($\sim10$~Myr) and distance 
($\sim1$~kpc) because of the cluster NGC\,1502 \citep{tap91,pau05}.  A possible 
abnormal interstellar reddening has been suggested, as diagnosed by the 
total-to-selective extinction, $R_V = A_V/E(B-V)$ from 2.42 \citep{tap91} to 
2.57 \citep{pan03}, as compared to the average $R_V=3.1$ in the diffuse 
interstellar medium.  A reduced value of $R_V$ is indicative of a smaller 
than typical average grain size, likely as the consequence of photoevaporation 
by OB stars in the region.  The variation of $R_V$ has little effect on our 
analysis of the properties of G144.
 
The cluster G144 therefore is at the same distance as the Cam\,OB1 association 
which makes it physically a part of that association.  The newly found OB 
stars in 1A (Figure~\ref{obcmd}, left panel) and 1B (Figure~\ref{obcmd}, right 
panel) considerably augment the OB star inventory in the region.  For G144 
(Figure~\ref{obcmd}, middle panel), most PM candidates are distributed along 
the ZAMS.  YSO candidates such as stars A, B, and D are located above the MS, 
i.e., at ages of $\lesssim1$~Myr and have NIR excess.  The existence of CTTSs 
sets a stringent age limit for the cluster---no more than a couple of 
megayears.

\begin{figure}
\plotone{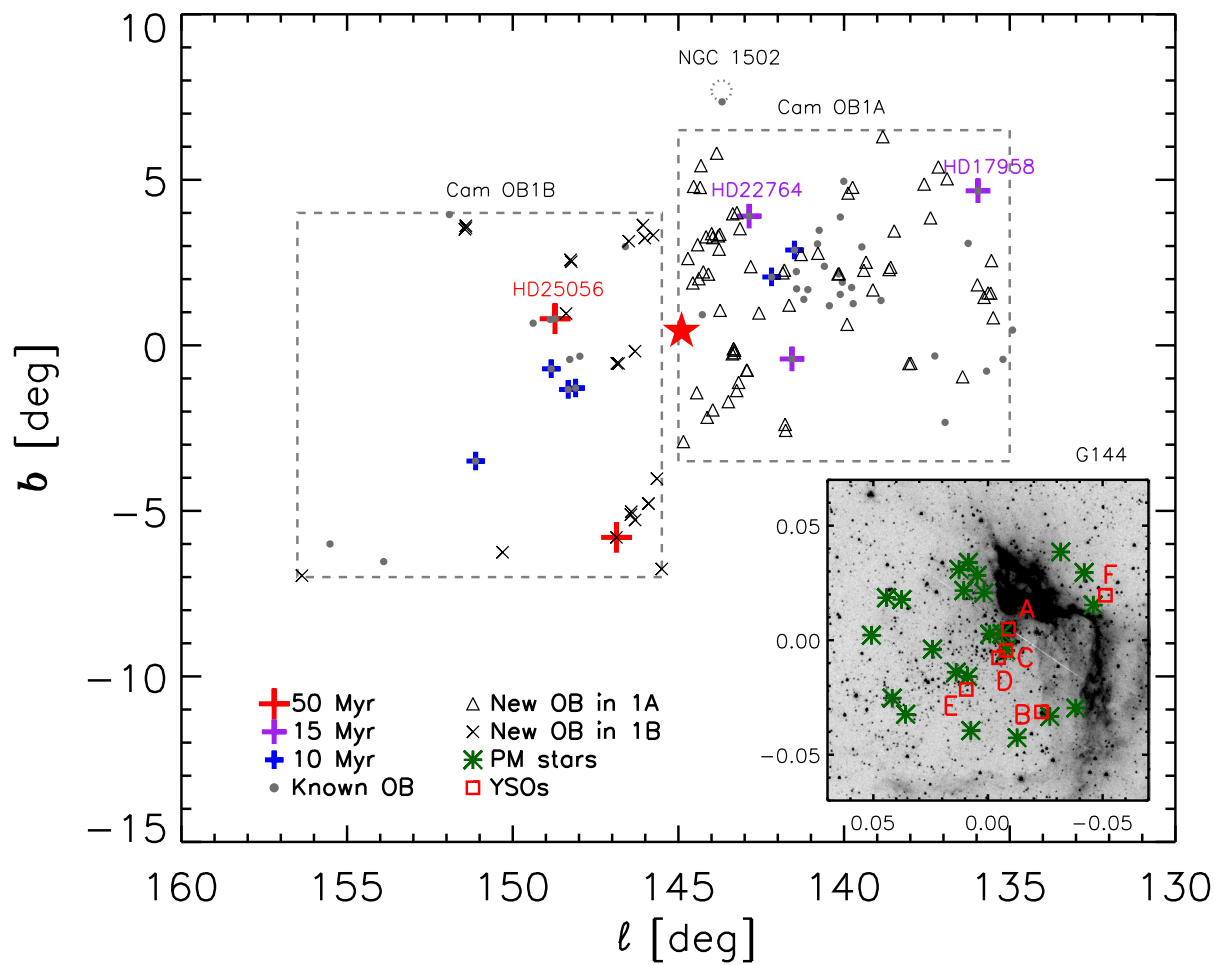}
\caption{  Atlas of the Cam\,OB1 complex and G144.  Gray circles are known OB 
members.  The dashed boxes mark the boundaries of Cam\,OB1A and 1B members.  
Black crosses and triangles mark newly found OB stars.  The position of G144 
is marked with a red star, whose expanded view, centered around 
$\ell\approx144\fdg9$ and $b\approx0\fdg4$, is shown as the inset to the 
bottom right.  Green asterisks are PM candidates and red boxes represent YSO 
candidates, which are labeled individually.  Members following isochrones of 
ages 50, 15, or 10 Myr are represented with pluses.  }
\label{atlas}
\end{figure}

Figure~\ref{atlas} illustrates the positional distribution of the young stellar 
population in Cam\,OB1.  Subgroup 1B is associated with a molecular cloud 
ring \citep{str08}.  It is conceivably a remnant bubble created by a supernova, 
devoid of gas and dust, and the newly found OB stars in 1B, like the known 
sample, are distributed outside the ring.  The OB stars in 1B are also 
collectively spread toward 1A.  The cloud complex in 1A and 1B has a 
filamentary shape \citep[see Figure 1 in ][]{str08}, along which the majority of 
the young stars are densely populated.  

It has been proposed that Cam\,OB1 has sustained star formation for the last 
100~Myr, with a stellar age progressively younger toward the northern part of 
the complex \citep{lyd01}.  A few evolved member candidates shed crucial light 
on the star formation history in the region.  In Figure~\ref{obcmd}, one sees a 
handful of candidates in 1A along the 10--15~Myr post-MS isochrones.  In 
particular, two very bright candidates, HD\,22764 (spectral type of K4\,Ib) and 
HD\,17958 (spectral type of K3\,Ib), could be as old as 15~Myr.  Both are 
well within the spatial (see Figure~\ref{atlas}) and PM (see 
Figure~\ref{ppmob}) groupings like other members.  Thus the age spread in 1A 
should be secure.  

In 1B, candidates considerably older.  One candidate suggests an age 
as old as 50~Myr.  This star, HD\,25056 (spectral type of G0\,Ib), has a 
marginal PM (near the boundary circle in Figure~\ref{ppmob}) and is relatively 
isolated in location (see Figure~\ref{atlas}); its membership, and hence its 
isochrone age, is less secure.  The subgroup 1B is at least up to 10--15~Myr 
old.  It is also possible that, from the distribution of young stars, 1B 
 formed as the consequence of the ``collect-and-collapse'' mechanism 
\citep{elm77,deh05, zav06} with which massive stars sweep ambient material via 
their photoionization pressure or supernova shocks, so dense clumps are 
collected in a shell collapse to form the next-generation stars.  What we 
witness in G144 is the result of the most recent star birth in this cloud 
complex.  

\section{Conclusions}

We used photometric, spectroscopic, and kinematic data to characterize the open 
cluster candidate G144 found in our OC search pipeline.  The cluster is found 
to be a part of the Cam\,OB1 association.  A total of 23 member candidates have 
been identified, some still in the pre-main sequence stage, which are 
distributed within an angular diameter of $\sim7\arcmin$ at a distance 
of 1~kpc, this corresponds to a linear scale of $\sim2$~pc across.  The cluster 
is no more than a couple of megayears old.  

Moreover, we identified a total of 91 additional OB star candidates in 
subgroups 1A and 1B on the basis of the PMs of 43 known OB members.  
Cam\,OB1 is associated with a molecular cloud complex and is shown to have 
undergone sustained star formation in subgroups 1A and 1B (and also 1C which 
is not studied here) for at least the last 10--15~Myr.  The open cluster G144 
appears to highlight the latest episode of sequential star formation in 
the region.

\acknowledgments

We acknowledge financial support from Taiwan's National Science Council via 
grants NSC101-2628-M-008-002 and NSC101-2628-M-008-001 which made this study 
possible. We thank the HCT staff of their 
assistance with acquiring the spectral data reported here during the observing 
run.  This publication makes use of data products from the Two 
Micron All Sky Survey and the {\it Wide-field Infrared Survey Explorer}, which 
is a joint project of the University of Massachusetts and the Infrared 
Processing and Analysis Center/California Institute of Technology, funded by 
the National Aeronautics and Space Administration and the National 
Science Foundation.  This work is also based on observations made with the 
{\it Spitzer Space Telescope}, which is operated by the Jet Propulsion 
Laboratory, California Institute of Technology under a contract with NASA.  
We also acknowledge use of data obtained as part of IPHAS carried out at the 
Isaac Newton Telescope (INT). The INT is operated on the island of La Palma by 
the Isaac Newton Group in the Spanish Observatorio del Roque de los Muchachos 
of the Instituto de Astrofisica de Canarias. All IPHAS data are processed by 
the Cambridge Astronomical Survey Unit, at the Institute of Astronomy in 
Cambridge. 
 

\end{CJK*}

\end{document}